\documentclass[aps,prb,reprint,showkeys,superscriptaddress,floatfix]{revtex4-1}
\usepackage{amsfonts}
\usepackage{amssymb}
\usepackage{amsmath}
\usepackage{hyperref}
\usepackage{graphicx}
\usepackage{bookmark}
\begin{document}
\title{Graphene Plasmonic Fractal Metamaterials for Broadband Photodetectors}
\author{Francesco De Nicola}
\email[E-mail: ]{francesco.denicola@iit.it}
\affiliation{Graphene Labs, Istituto Italiano di Tecnologia, Via Morego 30, 16163 Genova, Italy}
\author{Nikhil Santh Puthiya Purayil}
\affiliation{Graphene Labs, Istituto Italiano di Tecnologia, Via Morego 30, 16163 Genova, Italy}
\affiliation{Physics Department, Universit\'a degli studi di Genova, Via Dodecaneso 33, 16146 Genova, Italy}
\author{Vaidotas Mi\^seikis}
\affiliation{CNI@NEST, Istituto Italiano di Tecnologia, Piazza San Silvestro 12, 56127 Pisa, Italy}
\author{Davide Spirito}
\affiliation{Nanochemistry Department, Istituto Italiano di Tecnologia, Via Morego 30, 16163 Genova, Italy}
\author{Andrea Tomadin}
\affiliation{Physics Department, Universit\'a di Pisa, Largo Bruno Pontecorvo 3, 56127 Pisa, Italy}
\author{Camilla Coletti}
\affiliation{CNI@NEST, Istituto Italiano di Tecnologia, Piazza San Silvestro 12, 56127 Pisa, Italy}
\author{Marco Polini}
\affiliation{Graphene Labs, Istituto Italiano di Tecnologia, Via Morego 30, 16163 Genova, Italy}
\author{Roman Krahne}
\affiliation{Nanochemistry Department, Istituto Italiano di Tecnologia, Via Morego 30, 16163 Genova, Italy}
\author{Vittorio Pellegrini}
\affiliation{Graphene Labs, Istituto Italiano di Tecnologia, Via Morego 30, 16163 Genova, Italy}
\date{\today}
\maketitle
\indent \textbf{Metamaterials have recently established a new paradigm for enhanced light absorption in state-of-the-art photodetectors. Here, we demonstrate broadband, highly efficient, polarization-insensitive, and gate-tunable photodetection at room temperature in a novel metadevice based on gold/graphene Sierpinski carpet plasmonic fractals. We observed an unprecedented internal quantum efficiency up to 100\% from the near-infrared to the visible range with an upper bound of optical detectivity of $\mathbf{10^{11}}$ Jones and a gain up to $\mathbf{10^{6}}$, which is a fingerprint of multiple hot carriers photogenerated in graphene. Also, we show a 100-fold enhanced photodetection due to highly focused (up to a record factor of $\mathbf{|E/E_{0}|\approx20}$ for graphene) electromagnetic fields induced by electrically tunable multimodal plasmons, spatially localized in self-similar fashion on the metasurface. Our findings give direct insight into the physical processes governing graphene plasmonic fractal metamaterials. The proposed structure represents a promising route for the realization of a broadband, compact, and active platform for future optoelectronic devices including multiband bio/chemical and light sensors.}\\
\indent Nowadays, the conversion of light into electrical signals is at the heart of several technologies. Notwithstanding the high level of maturity, the large scale, and diversity of application areas, the need for a compact photodetection platform with higher performance in terms of speed, efficiency, spectral range, as well as flexibility, semitransparency, and compatibility with complementary metal-oxide-semiconductor (CMOS), is becoming more eminent\cite{Koppens2014}.\\ 
\indent Such features can be greatly fulfilled by combining functional systems with metamaterials \cite{Zheludev2012} (i.e. structured materials on the subwavelength scale with engineered electromagnetic properties). This leads to the concept of metadevices \cite{Zheludev2012}, a logical extension of the metamaterial paradigm, where interactions are nonlinear and the response is dynamic, using systems with spatially variable elements. Generally, such metadevices can operate at low voltages, which is a competitive advantage over conventional technology exploiting bulk and expensive electro-optical crystals, often non-integrated in silicon photonics. However, the bandwidth of such metamaterials is generally narrow due to their resonant nature and most of them are working in the infrared band. Mastering control in the optical regime is technologically a more challenging task, due to the expensive and demanding fabrication of nanostructures, resulting from the need of high-resolution electron- or ion-beam techniques. Also, devices working in a wide spectral range are important for multiband photodetection \cite{Liu2011}, on-chip sensing of multiple analytes \cite{Aslan2016}, as well as multiplexed fluorescence and Raman detection \cite{Aouani2013}. Hybridization of low-dimensional carbon allotropes with plasmonic metamaterials \cite{Aygar2016,Fang2017,Nikolaenko2010,Yao2014a,Yao2014,Zhu2013a,Lee2012,Fang2012,Echtermeyer2011,Papasimakis2010,Emani2014} is known to provide broadband and strong light-matter interactions, improvement of the nonlinear response due to the electromagnetic field enhancement given by the metamaterial, and compatibility with well-established silicon photonic and CMOS technologies \cite{Gan2013,Liu2011a}, therefore realizing a platform for high-speed \cite{Mueller2010} light modulation and detection on the same chip.\\
\indent In this regard, graphene --- a two-dimensional system composed of carbon atoms arranged in a hexagonal lattice --- has rapidly established itself as intriguing building block for optoelectronic applications \cite{Bonaccorso2010}, with a strong focus on various photodetection frameworks \cite{Koppens2014}. The versatility of this exceptional material enables its application in a number of areas including ultrafast and electrically tunable detection of light over a wide spectral range from UV to THz \cite{Liu2011a,Massicotte2016,Ding2020,Tielrooij2015,Yao2014,Yao2014a,Liu2014,Freitag2013,Gan2013,Freitag2013a,Konstantatos2012,Vicarelli2012,Liu2011,Gabor2011,Mueller2010,Xia2009,Fang2012,Echtermeyer2011}. However, single-layer graphene absorbs 2.3\% of the incident light in a large spectral range \cite{Li2008}, which is remarkably high for an atomically thin material, but it leads to a low photocarrier absorption efficiency that constitute a limiting factor for highly sensitive state-of-the-art photodetectors. By coupling plasmonic metamaterials with graphene, localized surface plasmons \cite{Maier2007} (LSP) --- collective charge oscillations at the interface between a subwavelength metal object and a dielectric --- can be excited in photodetectors to improve graphene absorption by the resulting field enhancement, therefore increasing the device responsivity. However, plasmonic structures with periodic geometries can be used only for filtering light polarization and selective absorption at designed frequencies \cite{Yao2014,Liu2011,Yao2014a,Echtermeyer2011,Gilbertson2015}, restricting their application field. Recently, the first graphene plasmonic fractals \cite{Fang2017,Aygar2016} have been realized, reporting a polarization-insensitive, gate-tunable, and enhanced photodetection. So far, such metamaterials are still narrow-band and have low photodetection efficiency with poor electrical modulation.\\
\indent Here, we show that plasmonic fractal metamaterials can be designed to significantly improve broadband and polarization-independent light absorption in graphene photodetectors, thus enabling multiband photodetection. A novel graphene-based electrically active metadevice for efficient control and detection of radiation in the visible (VIS)--mid-infrared (MIR) range was realized. The metamaterial consists of an Au Sierpinski carpet (SC) fractal \cite{DeNicola2018} fabricated on a graphene/SiO$_2$/Si substrate. The gold/graphene (Au/G) fractal absorbs light 25 times stronger than an Au SC. As a result of the strong electromagnetic field enhancement (up to $|E/E_{0}|\approx20$) within the interstructure gaps, the graphene-light interaction is greatly enhanced (up to an unprecedented factor of $|E/E_{0}|^{4}\approx10^{5}$) at designed frequencies. This provides a multimodal plasmonic response that covers a broad range of the electromagnetic spectrum (VIS--MIR). Owing to the continuous nature of the Au/G fractal metamaterial, both its absorption intensity and bandwidth can be controlled electrically by varying the charge carrier concentration in graphene, through a gate voltage. This approach allows modulation of the device optical reflectance up to 23\% in intensity and 60\% in photon energy, much higher than that reported by Aygar et al. \cite{Aygar2016}. We observed in the optical range up to 100\% internal quantum efficiency, remarkably higher than that reported by Fang et al. \cite{Fang2017}, with a gain up to $10^{6}$ and an upper bound of optical detectivity of $2\times10^{11}$ cm Hz$^{1/2}$W$^{-1}$. Moreover, we present up to 100-fold enhancement of the responsivity, with respect to an unpatterned graphene photodetector.\\
\begin{figure*}[ht]
\centering
\includegraphics[width=0.70\textwidth]{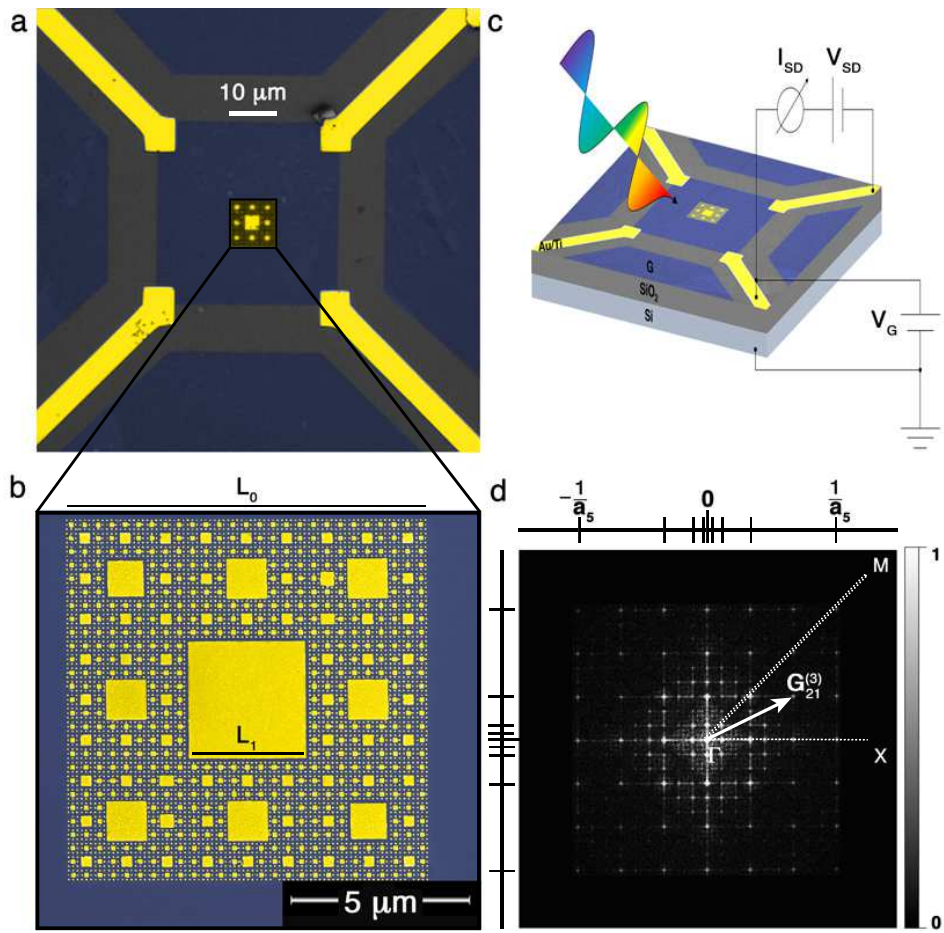}
\caption{Au/G plasmonic fractals. \textbf{a}, False color optical microscopy micrograph of a $35\pm3$ nm thick Au SC for $t=5$ order (in yellow) deposited on a CVD-grown, single-layer, single-crystal graphene (in blue) transferred on a Si/SiO$_{2}$ substrate (in gray). Four leads of the photodetector device are also shown (in yellow). \textbf{b}, Scanning electron microscopy micrograph of the Au/G SC. \textbf{c}, Scheme of the Au/G photodetector realized by Blender software, www.blender.org. \textbf{d}, Fast Fourier transform of the SEM micrograph in \textbf{b}. The $\Delta=\overline{\Gamma X}$ and $\Sigma=\overline{\Gamma M}$ directions in the fractal reciprocal lattice are marked along with the first pseudo-Brillouin zones $[-\pi/a_{t},\pi/a_{t}]^{2}$ of the different fractal orders. Starting from the origin, each point of the SC reciprocal lattice is defined by a reciprocal lattice vector $G_{ij}^{(t)}=2\pi\sqrt{i^{2}+j^{2}}/a_{t}$.}
\label{fig:Figure1}
\end{figure*}
\begin{figure*}[ht]
\centering
\includegraphics[width=1\textwidth]{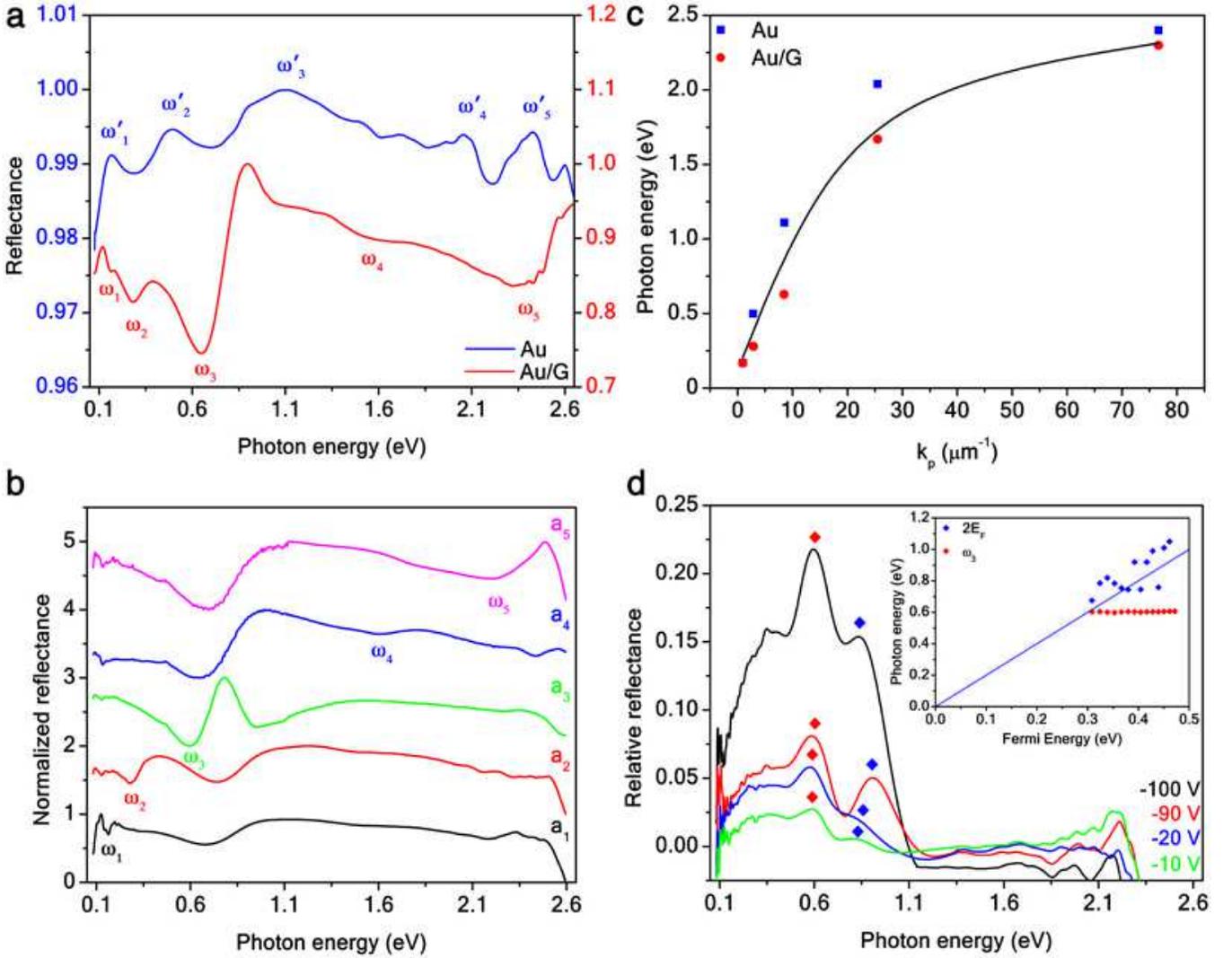}
\caption{Au/G fractal optical properties. \textbf{a}, Experimental reflectance $R=r/r_{sub}$ spectra of $35\pm3$ nm thick Au (blue solid line) and Au/G (red solid line) SCs for $t=5$ order. Localized surface plasmon modes $\omega^{\prime}_{1}=0.17$ eV, $\omega^{\prime}_{2}=0.50$ eV, $\omega^{\prime}_{3}=1.11$ eV, $\omega^{\prime}_{4}=2.04$ eV, and $\omega^{\prime}_{5}=2.40$ eV for the Au and $\omega_{1}=0.16$ eV, $\omega_{2}=0.28$ eV, $\omega_{3}=0.63$ eV, $\omega_{4}=1.67$ eV, and $\omega_{5}=2.30$ eV for the Au/G SCs are marked. \textbf{b}, Experimental normalized reflectance $R=r/r_{sub}$ of $35\pm3$ nm thick Au/G periodic arrays with lattice constants $a_{t}$ and LSP modes $\omega_{1}=0.16$ eV, $\omega_{2}=0.28$ eV, $\omega_{3}=0.59$ eV, $\omega_{4}=1.59$ eV, and $\omega_{5}=2.20$ eV. Curves are offset by 1. \textbf{c}, Dispersion relations of the LSP modes as a function of the plasmon wavevector for Au (blue squares), Au/G periodic arrays (red dots). Black curve is the result of electromagnetic calculations. \textbf{d}, Experimental relative reflectance $\left(R-R_{CNP}\right)/R_{CNP}$ of the Au/G SC at $t=5$ order for different gate voltages. The LSP mode $\omega_{3}$ (red diamonds) and the interband absorption threshold $2E_{F}$ (blue diamonds) are marked. \textbf{d, inset}, Dispersion relations of the interband absorption threshold $2E_{F}$ (blue diamonds) and the Au/G SC $\omega_{3}$ mode (red diamonds) as a function of the graphene Fermi level. The blue solid line is the law $\hbar\omega=2E_{F}$.}
\label{fig:Figure2}
\end{figure*}
\begin{figure*}[ht]
\centering
\includegraphics[width=1\textwidth]{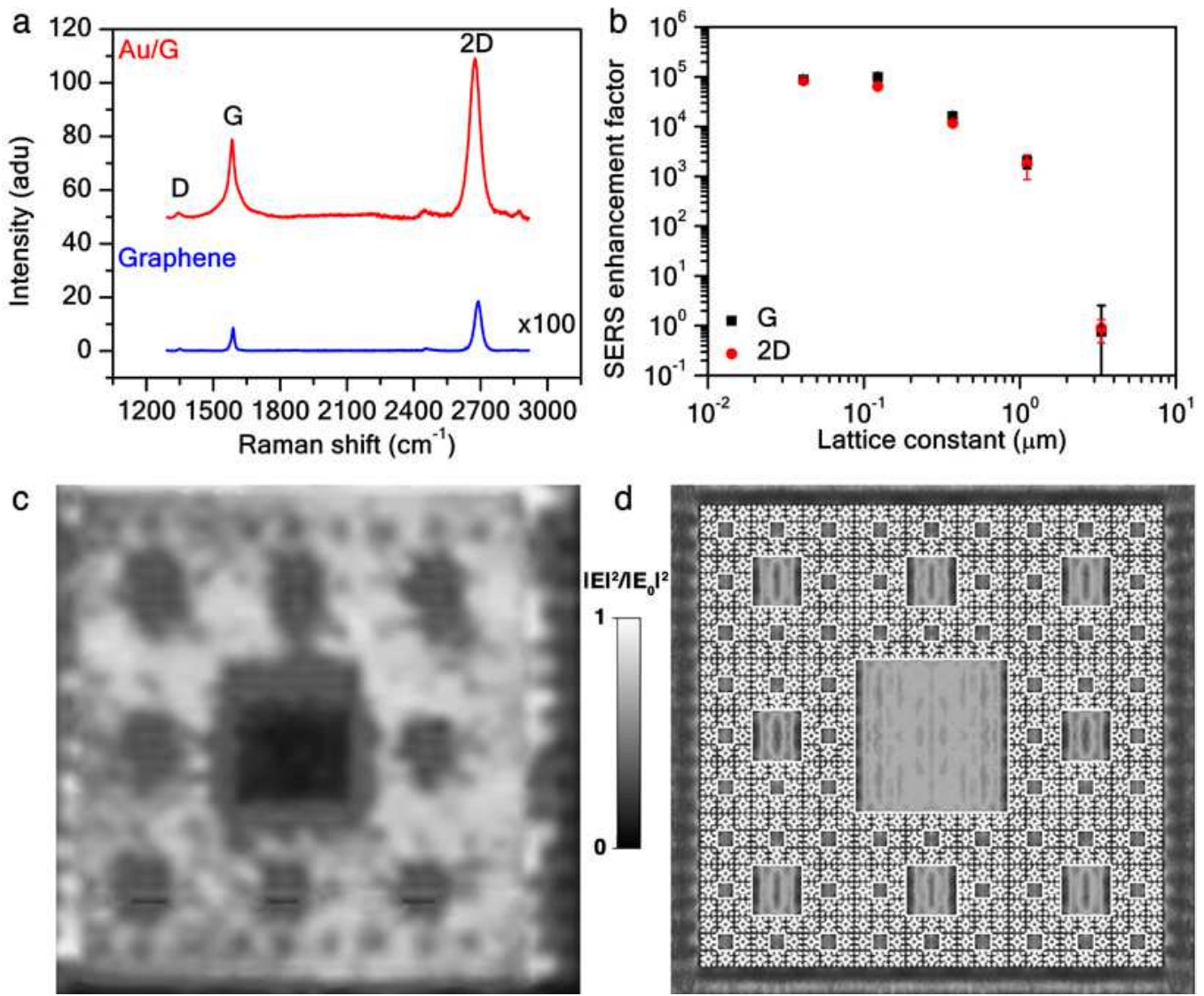}
\caption{Au/G fractal surface electromagnetic enhancement. \textbf{a}, Averaged Raman spectra at $\lambda_{ex}=514$ nm of the Au/G SC for $t=5$ order and for a reference unpatterned graphene. Curves are offset by 50. The graphene D, G, and 2D vibrational bands are shown at 1343-1349 cm$^{-1}$, 1584-1589 cm$^{-1}$, and 2674-2689 cm$^{-1}$, respectively. \textbf{b}, SERS enhancement of the graphene G (black squares) and 2D (red dots) bands at $\lambda_{ex}=514$ nm as a function of the SC lattice constant $a_{t}$, in log-log scale. Experimental electric field enhancement $\vert E\vert^{2}/\vert E_{0}\vert^{2}$ maps of the BCB vibrational band $\omega^{\star}=1655$ cm$^{-1}$ at $\lambda_{ex}=532$ nm for the $t=5$ order SC (\textbf{c}) and its electromagnetic simulation obtained by CST Microwave Studio, www.cst.com (\textbf{d}). Each map is normalized to its maximum value. Note that since the laser spot diameter is 1 $\mu$m, the instrument averages out over $L_{4}$ and $L_{5}$ structures.}
\label{fig:Figure3}
\end{figure*}
\begin{figure*}[ht]
\centering
\includegraphics[width=1\textwidth]{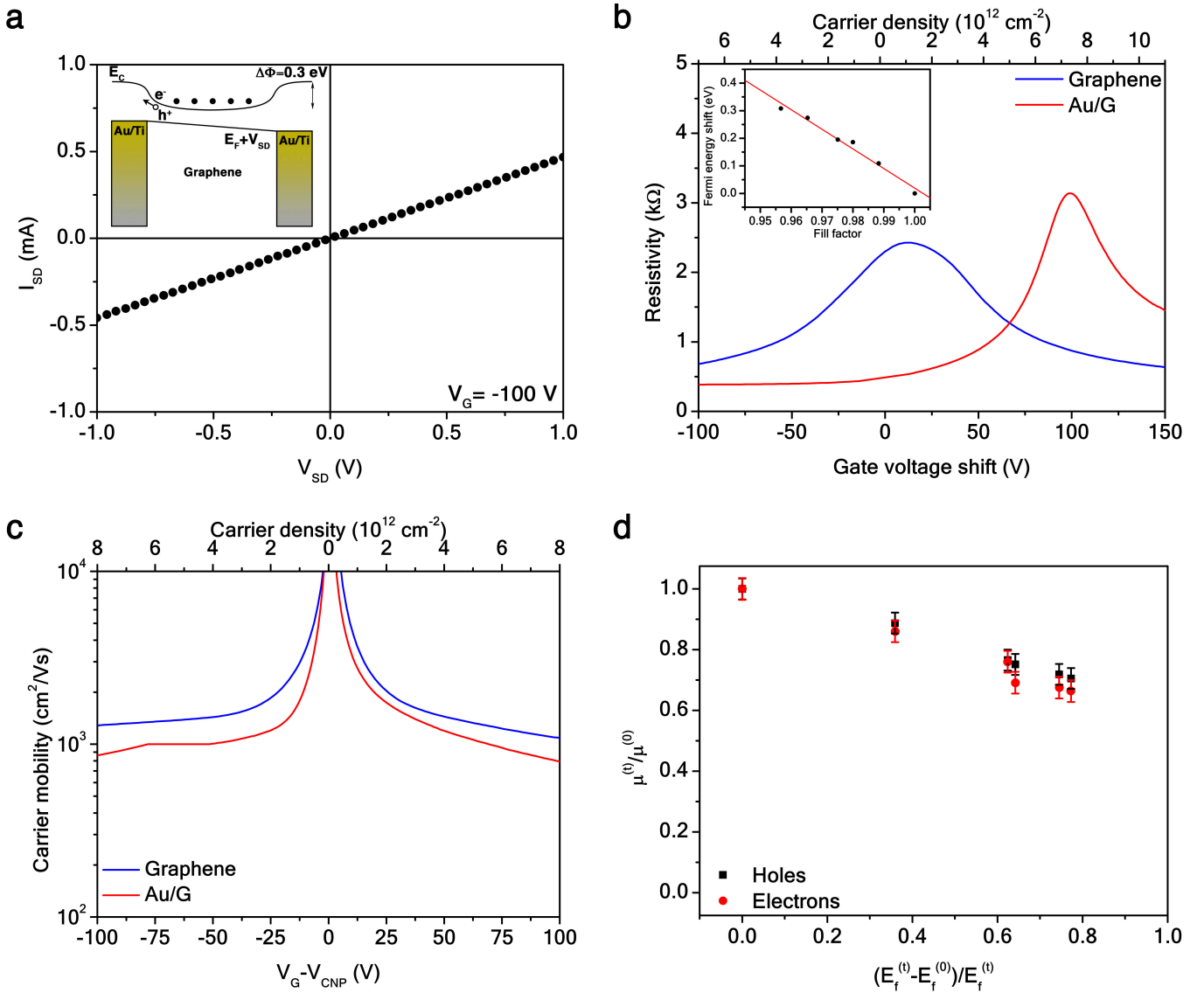}
\caption{Au/G fractal electrical properties. \textbf{a}, $I_{SD}$-$V_{SD}$ characteristics at $V_{G}=-100$ V of the Au/G SC for $t=5$ order. \textbf{a, inset}, Scheme of the conduction band $E_{c}$ spatial profile of the Au/G devices between two Au electrodes, under a positive bias $V_{SD}$. $E_{F}$ is the junction Fermi level at equilibrium and $\Delta\Phi$ the contact potential. \textbf{b}, Resistivity at $V_{SD}=0.1$ V of the Au/G SC for $t=5$ order and a reference unpatterned graphene as a function of the carrier density and the gate voltage shift with respect to the unpatterned graphene CNP. \textbf{b, inset}, Fermi level shift as a function of the Au/G fractal fill factor. The red solid line is a linear fit of the data. \textbf{c}, Carrier mobility at $V_{SD}=0.1$ V of the Au/G SC for $t=5$ order and a reference unpatterned graphene as a function of the  carrier density and the relative gate voltage to the unpatterned graphene CNP. \textbf{d}, Normalized mobility at a carrier density $n=6\cdot10^{12}$ cm$^{-2}$ of Au/G fractals for $t=0$-5 to the value for $t=0$ as a function of the relative Fermi level shift to the value for $t=0$. Error bars are due to averages over multiple resistivity measurements on the same sample.}
\label{fig:Figure4}
\end{figure*}
\begin{figure}[ht]
\centering
\includegraphics[width=0.5\textwidth]{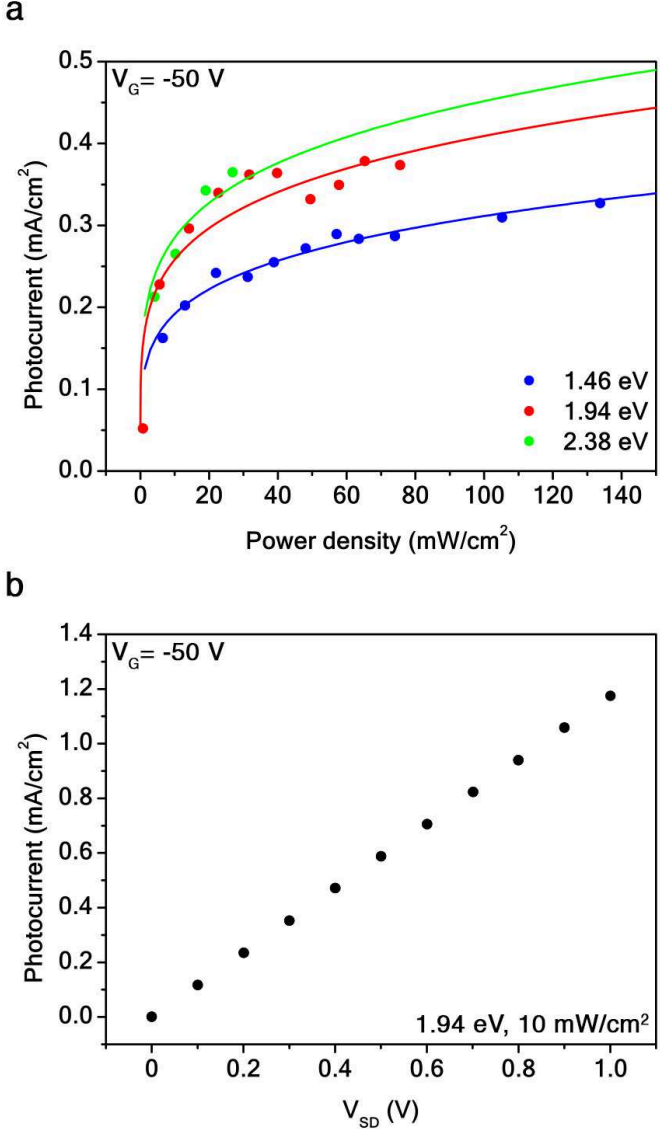}
\caption{Au/G $t=5$ order fractal photodetectors. \textbf{a}, Photocurrent at $V_{G}=-50$ V and $V_{SD}=0.1$ V as a function of the power of 1.46 eV (blue dots), 1.94 eV (red dots), and 2.38 eV (green dots) lasers. Solid lines are power law fits $J\propto P^{\alpha}$, with $0<\alpha<1$, for $P>6$ mW/cm$^{2}$. The fits return $\alpha=0.2$. \textbf{b}, Photocurrent at $V_{G}=-50$ V as a function of the $V_{SD}$ bias for a 1.94 eV laser of power 10 mW/cm$^{2}$.}
\label{fig:Figure5}
\end{figure}
\begin{figure*}[ht]
\centering
\includegraphics[width=1\textwidth]{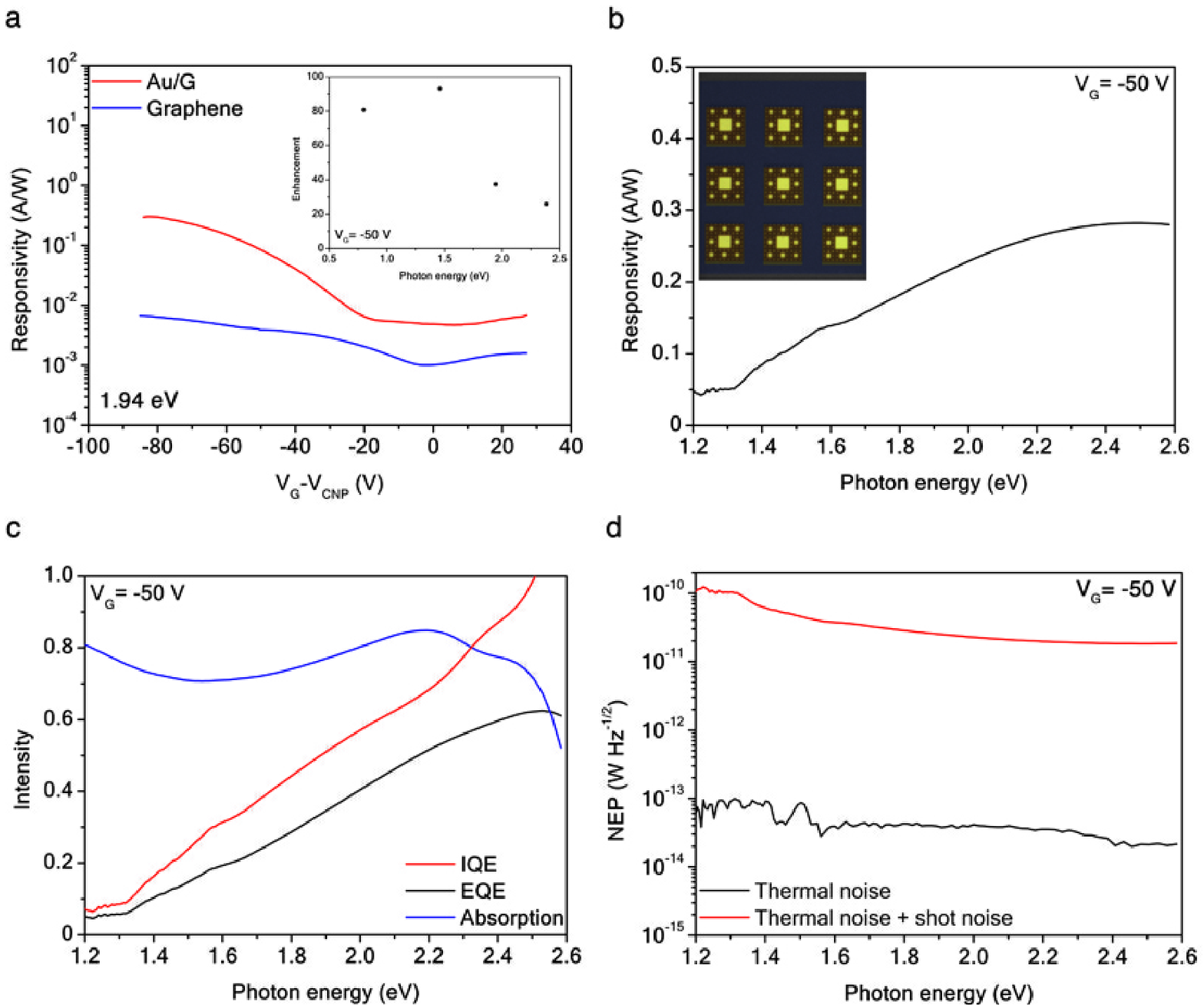}
\caption{Au/G $t=5$ order fractal array photodetector efficiencies. \textbf{a}, Responsivity of an Au/G SC array and an unpatterned graphene photodetector at $V_{SD}=0.1$ V as a function of the gate voltage. \textbf{a, inset}, Responsivity enhancement factor of an Au/G SC array with respect to an unpatterned graphene photodetector ($t=0$) at $V_{G}=-50$ V and $V_{SD}=0.1$ V. \textbf{b}, Responsivity of an Au/G SC photodetector as a function of the photon energy of a panchromatic source at $V_{G}=-50$ V. \textbf{b, inset} False color optical microscopy micrograph of the $3\times3$ array on the device channel. \textbf{c}, Au/G SC device total absorption ($A=1-r$), internal and external quantum efficiencies at $V_{G}=-50$ V and $V_{SD}=0.1$ V. \textbf{d}, Total NEP (red solid curve) and only thermal contribution (solid black curve) at $V_{SD}=0.1$ V as a function of a panchromatic source at $V_{G}=-50$ V in linear-log scale.}
\label{fig:Figure6}
\end{figure*}
\section*{Results and Discussion}
\label{results}
\indent Deterministic fractals \cite{Mandelbrot1982} are self-similar objects generated by geometrical rules, having a non-integer (Hausdorff-Besicovitch) dimension. Sierpinski carpets can be generated by a recursive geometrical algorithm employing a Lindenmayer system implementation \cite{DeNicola2018}. We designed Au/G SCs starting from a graphene unit cell of side $L_{0}=10$ $\mu$m that is divided into a $3\times3$ array of sub-cells of lateral size $L_{1}=L_{0}3^{-1}$, with an Au square placed in the central sub-cell. By iteratively applying the same rule to the generated sub-cells of size $L_{t}=L_{0}3^{-t}=L_{0}\mathcal{L}_{t}$, it is possible to obtain fractals for higher orders $t$ of complexity \cite{DeNicola2018}. In this way, we indirectly patterned graphene, thereby realizing an Au/G fractal metamaterial that can be thought as a discrete Au SC and a complementary continuous graphene SC. At each iteration the side of the sub-cells is reduced by a factor $\mathcal{L}=3^{-1}$. Since the number of graphene sub-cells in the SC increases by a factor $\mathcal{N}=8$ at each iteration, the fractal dimension is $d_{H}=\log{\mathcal{N}}/\log{\mathcal{L}^{-1}}\approx1.89$ \cite{DeNicola2018}.\\
\indent By employing electron-beam lithography, reactive-ion etching, metal evaporation, and lift-off techniques (see Methods), we fabricated the first five orders of an Au/G SC on a Si/SiO$_{2}$ substrate (see Supplementary Information 1). The nominal thickness of the Au squares is $35\pm3$ nm, while their lateral sizes are $L_{1}=3.33\pm0.05$ $\mu$m, $L_{2}=1.11\pm0.01$ $\mu$m, $L_{3}=370\pm17$ nm, $L_{4}=123\pm7$ nm, and $L_{5}=41\pm3$ nm. A representative optical micrograph of the sample for $t=5$ order is shown in Figure \ref{fig:Figure1}a along with a scanning electron microscopy (SEM) micrograph in Figure \ref{fig:Figure1}b, which illustrates the excellent homogeneity and uniformity of the fractal structure. In Figure \ref{fig:Figure1}c, a scheme of an Au/G fractal photodetector is depicted. In order to understand the physics governing the metadevice, before presenting the device characterization we discuss the optical properties of the Au/G fractal metamaterial in the following.\\
\indent As we have previously reported in Ref. \onlinecite{DeNicola2018}, the Au SC can be regarded as a metallo-dielectric photonic quasi-crystal. In Figure \ref{fig:Figure1}d the singular-continuous \cite{Negro2012} reciprocal lattice of the SC for $t=5$ order is shown, by computing the fast Fourier transform of its SEM micrograph. Such a reciprocal lattice is self-similar as its direct counterpart \cite{DeNicola2018}. It follows that the diffraction patterns and the related optical spectra of such fractals are themselves self-similar \cite{Negro2012,DeNicola2018}. Also, the SC has a larger number of points in the reciprocal space than a periodic array, since its fractal reciprocal lattice is a superposition of five periodic lattices with different constants $a_{t}=3L_{t}$, under the periodic approximation \cite{Negro2012,Gambaudo2014,DeNicola2018}.\\
\indent In Figure \ref{fig:Figure2}a, a comparison between the reflectance $r$ of an Au and Au/G SC divided by the reflectance $r_{sub}$ of the SiO$_{2}$/Si substrate is shown. The reflectance of the substrate is dominated by a minimum around 0.7 eV due to thin-film interference effects between SiO$_{2}$ and Si. The spectra of Au and Au/G SCs exhibit multiple features in the VIS--MIR range, which for the Au SC appear as maxima $\omega^{\prime}_{n}$ due to the radiative scattering in air, while for the Au/G SC result in shifted minima $\omega_{n}$ owing to the thin-film interference effects between Au and graphene. In Ref. \onlinecite{DeNicola2018}, we have reported that the optical spectrum of an Au SC at a given fractal order $t$ mainly manifests $t$ diffraction-mediated LSP resonances $\omega_{n}\approx2\pi c/3L_{0}\mathcal{L}_{n}$, where $n=1,2,\ldots t$, for small LSP wavevectors $k_{p}$. For large $k_{p}$, $\omega_{n}\equiv\omega_{sp}/\sqrt{3}=2.40$ eV with $\omega_{sp}=4.17$ eV the Au surface plasmon \cite{Maier2007} (see Supplementary Information 2). These LSP resonances are self-similar as their photon energy has the same scale-invariance law ($\mathcal{L}_{t}=8^{-t/d_{H}}$) as the SC lattice constant $a_{t}$, depending on the fractal dimension $d_{H}$ (see Supplementary Information 2). On the other hand, graphene surface plasmons polaritons are expected at lower photon energy \cite{Tantiwanichapan2017}. Furthermore, the relative reflectance change $(R_{Au}-R_{Au/G})/R_{Au}$ between the two samples is up to 25\% at $\omega_{3}$, which means a remarkable relative absorption variation $(R_{Au/G}-R_{Au})/(1-R_{Au})$ of 2500\%. Therefore, graphene can be exploited as an ultrathin, broadband, anti-reflective coating. From the comparison with the reflectance spectra of Au/G periodic arrays in Figure \ref{fig:Figure2}b, it is possible to observe that the energy of the Au/G SC 
resonances scales with the periodicity $a_{t}$ of the Au structures. Also, the SC resonances in Figure \ref{fig:Figure2}a are blue-shifted with respect to those of the periodic arrays in Figure \ref{fig:Figure2}b, due to a far-field hybridization of LSPs belonging to the different fractal orders \cite{DeNicola2018}. In particular, the $\omega_{3}$ mode overlaps with the SiO$_{2}$ minimum. From data in Figure \ref{fig:Figure2}b, we can draw the dispersion relation of the hybrid Au/G LSP modes in Figure \ref{fig:Figure2}c. From the electromagnetic calculations (see Supplementary Information 3), $k_{p}=2\pi/L_{t}=6\pi/a_{t}$ is the self-similar LSP wavevector of the Au SC. From the plot we observe that the out-of-plane coupling between the Au SC LSP modes and graphene is two times stronger than the in-plane coupling between the LSP modes of the Au SC squares (see Supplementary Information 2).\\
\indent Moreover, the Au SC interaction with graphene allows to tune the Au/G LSP modes by applying a gate voltage to the device. The intensity of the relative reflectance to the charge neutrality point (CNP), $\left(R-R_{CNP}\right)/R_{CNP}$, for the five LSPs can be enhanced up to 23\% (\ref{fig:Figure2}d). Also, we noticed that only the position in energy of the $2E_{F}$ graphene interband threshold (Figure \ref{fig:Figure2}d inset) can be modulated up to 60\%, following the law $2E_{F}=2\hbar v_{F}\sqrt{\pi n}$, being $n\approx C\left(V_{G}-V_{CNP}\right)$ the charge carrier density, $C=\varepsilon_{SiO_{2}}\varepsilon_{0}/et_{ox}=7.2\times10^{10}$ V$^{-1}$cm$^{-2}$ the SiO$_{2}$ gate capacitance per area, $V_{G}$ the gate voltage, $V_{CNP}=30$ V the voltage at the charge neutrality point, and $v_{F}=10^{8}$ cm/s the graphene Fermi velocity. We did not observe any shift in energy for the the other LSP modes for such values of applied gate voltage. The dispersion of $\omega_{3}$ is shown in Figure \ref{fig:Figure2}d, for instance.\\
\indent Additionally, in Ref. \onlinecite{DeNicola2018} we demonstrated that having a centrosymmetric geometry, plasmonic SCs are insensitive to the linear polarization of the incident light (i.e., the SC response does not change if light polarization is parallel to the edge of the SC or to its diagonal). This is an important feature in order to realize polarization-independent devices \cite{Fang2017}, since other aperiodic metamaterial \cite{Emani2014,Papasimakis2010,Lee2012,Aygar2016} and periodic metamaterials \cite{Echtermeyer2011,Yao2014a,Yao2014,Liu2011} usually depend strongly on light polarization.\\
\indent Plasmonic fractals provide a hierarchical spatial distribution of near-field electromagnetic enhancement, due to LSP modes supported on the surface of the self-similar resonators forming the whole structure \cite{Negro2012,DeNicola2018}. When a mode of the Au fractal is excited, a considerable field enhancement not only near the squares but also across the graphene layer is observed. In fact, graphene represents a unique model system where nonlinear applications, such as surface enhanced Raman spectroscopy (SERS), can be studied in detail \cite{Schedin2010,Heeg2013,Zhang2014,Zhao2014,Iyer2014}. In Figure \ref{fig:Figure3}a, average Raman spectra for the $t=5$ order of the Au/G SC and for an unpatterned graphene sample $(t=0)$ are plotted. A SERS enhancement factor on the Au/G SC can be evaluated in Figure \ref{fig:Figure3}b as a function of the SC lattice constant $a_{t}$ for the graphene G and 2D phonon bands. As for Au SCs, the SERS enhancement factor follows a power law \cite{DeNicola2018}, but for Au/G SCs is 10-fold higher for the smallest feature, achieving a value of $10^{5}$. Additionally, as shown in Figure \ref{fig:Figure3}a, depositing metal structures on graphene represents an elegant way to pattern it without directly etching it, thus without altering the graphene crystalline structure (see Supplementary Information 4). Furthermore, by depositing on the $t=5$ order Au/G SC a thin layer of Brilliant Cresyl Blue (BCB) dye as a probe (see Supplementary Information 5), it is possible to map the SERS enhancement all over the fractal surface (Figure \ref{fig:Figure3}c). The contour plot presents a maximum of contrast among the squares with size $L_{4}$ and $L_{5}$ in excellent agreement with the electromagnetic simulation illustrated in Figure \ref{fig:Figure3}d (see Supplementary Information 3). At the probed excitation wavelength the dye is not resonant \cite{DeNicola2018}, therefore the observed signal is entirely related to a surface-enhanced electromagnetic effect.\\
\indent Recently, the quantum properties of electrons in fractals have been investigated both theoretically \cite{Veen2016} and experimentally \cite{Kempkes2018}, however their classical electronic behavior is has not been experimentally reported so far. As shown in Figure \ref{fig:Figure1}a, c, the phototransistor channel in the photodetector devices consists of a $40\times40$ $\mu$m CVD-grown, single-crystal single-layer graphene flake sensitized with an Au SC fractal (see Methods). Graphene is the active material of the device and the Au fractal acts as a plasmonic sensitizer to improve the overall light sensing. In the inset of Figure \ref{fig:Figure4}a, a scheme of the electronic band profile of the Au/G photodetector device is depicted. Upon contact, graphene forms an ohmic junction with the Au electrodes (Figure \ref{fig:Figure4}a). Therefore, an electrostatic contact potential $\Delta\Phi=0.3$ eV is built, due to the difference between the graphene and Au work functions $\Phi_{G}=4.5$ eV \cite{Bonaccorso2010} and $\Phi_{Au}=4.8$ eV \cite{Kittel2005}, respectively. For the same reason, the Au fractal also provides an hole charge-transfer to graphene (Figure \ref{fig:Figure4}b), resulting in a p-type doping. As illustrated in the inset of Figure \ref{fig:Figure4}b, such a doping is linear with the Au/G fill factor, i.e. the area fraction of graphene uncovered by Au (see Supplementary Information 1). This effect leads to an increase of the hole carrier density $n$ in graphene, therefore to a decrease of the carrier mobility $\mu_{e}=\sigma/ne$, as a function of the fractal order or fill factor (Figure \ref{fig:Figure4}c, d).\\
\indent Since all the leads are made of the same material, a symmetric potential that traps the photogenerated electrons occurs in the device (Figure \ref{fig:Figure4}a, inset). By applying a bias voltage $V_{SD}$, the photogenerated holes can be extracted, provided that the graphene Fermi level $E_{F}$ compensates the trapping potential by applying for instance a gate voltage $V_{G}$ (see Supplementary Information 6).\\
\indent In Figure \ref{fig:Figure5}a, it is possible to observe that for a $t=5$ order device, the photocurrent divided by the illuminated active area $J=I_{ph}/(4L_{0})^{2}$ exhibits a nonlinear dependence on the incident power of light $P$, due to electrostatic and thermal effects, trapping, and recombination occurring at high values of power \cite{Sze1969,Fang2017}. The photocurrent response results linear up to 6 mW/cm$^{2}$ for $V_{G}=-50$ V at a photon energy of 1.94 eV. Furthermore, by photogating effect, the device produces a DC photovoltage that is dependent on the incident photon energy and power (see Supplementary Information 6). Moreover, the Au/G SC can linearly detect light down to 2 nA for $V_{G}=-50$ V at 1.94 eV (Figure \ref{fig:Figure5}b).\\
\indent In Figure \ref{fig:Figure6}a, b, the broadband and electrically tunable responsivity $\mathcal{R}=J/P$ of a $t=5$ order Au/G SC $3\times3$ array can be observed as a function of the applied gate voltage. An enhancement factor $\mathcal{R}_{Au/G}/\mathcal{R}_{G}$ up to 100-fold of the Au/G SC device over an unpatterned graphene device (see Supplementary Information 6) is presented in the inset of Figure \ref{fig:Figure6}a. The array was introduced in order to increase the photocurrent signal-to-noise ratio. In addition, the device total absorption $A=1-r$, and the external $EQE=\mathcal{R}\hbar\omega/e$ and internal $IQE=EQE/A$ quantum efficiencies are evaluated in Figure \ref{fig:Figure6}c. It is worth noting that in the spectral range investigated, the Au/G SC photodetector IQE is up to 100\% in correspondence of the $\omega_{5}$ LSP resonance.\\
\indent From the application perspective, the relevant figure of merit to characterize a photodetector is the optical noise-equivalent-power $NEP=\sum_{i}N_{i}/\mathcal{R}$, which corresponds to the lowest detectable power in 1 Hz output bandwidth \cite{Sze1969}. Usually, the noise level of photodetectors in current-operating mode is due to the shot noise\cite{Sze1969} $N_{1}=\sqrt{2eI_{SD}}$ and to the thermal Johnson-Nyquist contribution \cite{Sze1969} $N_{2}=\sqrt{4k_{B}T\sigma}$, which ultimately limits the NEP. In Figure \ref{fig:Figure6}d NEP is shown as a function of the incident photon energy for $V_{G}=-50$ V. At room-temperature, we observed a minimum NEP of $2\times10^{-14}$ W Hz$^{-1/2}$ for a 2.38 eV photon energy, while $D^{\ast}\equiv L/NEP=2\times10^{11}$ cm Hz$^{1/2}$W$^{-1}$ or Jones represents the device maximum specific detectivity at the same photon energy. Such a high value of detectivity is compatible with state-of-the-art two-dimensional photodetectors \cite{Koppens2014,Xie2017}.\\
\indent Generally, five main phenomena compete in the operation of a graphene-based photodetector: the photovoltaic, the bolometric, the photothermoelectric, the photogating, and the plasma-wave-assisted effect \cite{Koppens2014}. In our case of study, we can reasonably exclude the latter mechanism, as it is likely to occur for THz radiation\cite{Vicarelli2012} ($\omega\tau\ll1$, with $\tau\approx0.1$ ps the graphene carrier scattering time in our samples). In a metal-graphene-metal configuration, in which graphene is contacted with metal electrodes as source and drain, a p-n junction occurs close to the contacts because of the difference in the work functions between the metal and graphene \cite{Echtermeyer2014}. Consequently, a charge transfer with a shift of the graphene Fermi level in the region below the metal pads occurs. The junction results in an internal electric field able to separate the light-induced electron-hole pairs (photovoltaic effect). Unless the contacts are made of different materials, the photocurrent produced at both the contacts will be of opposite polarity for symmetry reasons, resulting in zero net current at small gate voltages close to the CNP \cite{Mueller2010}. Usually, in such devices a light-induced photogating and photothermoelectric effect are also present at gate voltages close to the CNP, while at larger gate voltages the bolometric effect is typically dominant \cite{Koppens2014,Echtermeyer2014,Freitag2013a}. We observed the photovoltaic and photogating effects in the unpatterned graphene photodetectors (see Supplementary Information 6). In the Au/G SC devices, at gate voltages close to the CNP and for a symmetrical electronic band profile, the photovoltaic and bolometric effects are expected to play a minor role in comparison to the photothermoelectric and photogating effects \cite{Freitag2013a}, due to the plasmonic Au SC sensitizer (see Supplementary Information 6). On the other hand, at sufficiently larger gate voltages ($V_{G}<-40$ V), where the photothermoelectric and photogating effects become smaller, the bolometric effect should be dominant. Therefore, we propose the following explanation for the photogeneration mechanism in Au/G SC photodetectors. When light is shone on the metadevice, it is absorbed by Au fractal LSPs generating electron-hole pairs in the sensitizer (gating layer), which are transferred to graphene (channel layer). This results in a photocurrent and a photovoltage due to the change in electronic temperature between the Au SC and graphene regions, thus a variation of the carrier density $\Delta n=C\Delta V_{G}$ (see Supplementary Information 6). Since the device electronic bands are symmetric, the photogenerated electrons are trapped in graphene for $\tau_{trap}\equiv\Delta n\hbar\omega/IQE P\approx27$ ms (see Supplementary Information 6). Therefore, only holes at relatively high gate voltages contribute to the photocurrent. For the $t=5$ order Au/G SC array at $V_{G}=-50$ V the hole carrier transit time is $\tau_{trans}\equiv L^{2}/\mu_{e}V_{SD}\approx20$ ns, with FET channel length $L=40$ $\mu$m and carrier mobility $\mu_{e}\approx8\times10^{3}$ cm$^{2}$V$^{-1}$s$^{-1}$. Therefore, the lower bound (i.e. by neglecting the circuit RC constant) of the device bandwidth is $\Delta f_{-3dB}\approx3.5/2\pi\tau_{trans}=30$ MHz, while the photoconductive gain is up to $G\equiv\tau_{trap}/\tau_{trans}=10^{6}$, providing evidence of a multiple carrier generation \cite{Gabor2011,Tielrooij2013} in graphene (see Supplementary Information 6). Evidently, such a high gain, produced by prolonged carrier lifetime due to trapping, limits the photodetector response speed and bandwidth.\\
\indent In summary, we presented a novel graphene-based, plasmonic metamaterial having a multiresonant optical response hierarchically distributed on the fractal structure, which can be employed to realize a highly sensitive and polarization-independent, broadband photodetector. The metadevice is compact and can operate at room temperature under small voltages, with low power consumption. The detectivity could be further improved by depositing the Au fractal on high-mobility graphene-based van der Waals heterostructures, such as G/hBN \cite{Banszerus2015}. Moreover, the deterministic chemical vapor deposition (CVD) growth of large area graphene\cite{Miseikis2017} would grant integration over wafer scale at a low cost of graphene-based photodetectors. Finally, the wide-range and high-gain spectral response of the Au/G fractal and its ability to hierarchically localize strong electromagnetic fields on a sub-wavelength scale could pave the way to a new generation of materials for integrated optics and optoelectronics, such as multiband photosensor pixel arrays for imaging cameras and optical communication devices.
\section*{Methods}
\label{sec:methods}
\subsection*{Device fabrication}
\label{sec:device}
\indent Single-layer, single-crystal graphene was grown on copper foil by CVD using an Aixtron BM cold-wall reactor at a temperature of 1060 $^{\circ}$C and a pressure of 25 mbar. A semi-dry transfer approach was used to transfer the graphene from Cu to the target substrate \cite{Miseikis2017}. The foil was spin-coated by a PMMA membrane and the graphene was electrochemically delaminated from the growth substrate in 1M NaOH solution. Using a frame, the graphene/PMMA membrane was removed from the electrolyte, rinsed in deionized water and dried in ambient conditions. The membrane was deposited on the target Si/SiO$_{2}$ substrates at 120 $^{\circ}$C and the PMMA carrier polymer was removed in acetone. Graphene and Au SCs were patterned by EBL (Raith 150-Two) on Si wafers (1-5 m$\Omega$ cm) with a $t_{ox}=285$ nm tick layer of thermal oxide. A layer of approximately 160 nm of 950K poly-methyl-methacrylate (PMMA A2 1:1, 1\% in anisole) was deposited by spincoating on the substrates, which were then post-baked for 7 min on a hot plate at 180$^{\circ}$ C. The electron beam was operated at 20 kV with a current of 35 pA. After exposure, graphene was etched by O$_{2}$ (5 sccm) reactive ion etching (Sentech instruments ICP-RIE 500) for 30 s at a RF power of 35 W and temperature of $15^{\circ}$ C in an Ar inert atmosphere (80 sccm). The SC patterns were exposed with a varying dose inversely proportional to the size of the squares of the fractal. Contacts pads and probes were exposed for the electrical measurements. The resist was developed for 30 s in a cold (8$^{\circ}$ C) 1:3 mixture of MIBK:IPA. A 35/5 nm thick Au/Ti film was deposited on the substrates at a rate of 0.2 \AA/s by electron-beam evaporation (Kenosistec) at an operating pressure of $10^{-6}$ mbar. Lift-off was then performed by hot ($80^{\circ}$ C) acetone. A wire bonder (F\&S Bontec Series 56) was used to wire the contact pads to an external circuit board. For SERS measurements, SCs were dip-coated for 1 h in 1 mM BCB [(C$_{17}$H$_{20}$ClN$_{3}$O)$_{2}$ZnCl$_{2}$] aqueous solution, then rinsed in deionized water to wash the excess molecules in order to form a thin layer $12\pm2$ nm thick, and finally dried in nitrogen flow.
\subsection*{Sample characterization}
\label{sec:characterization}
\indent Samples were characterized as in Ref. \onlinecite{DeNicola2018}. Scanning electron microscopy micrographs of Au/G SC samples deposited on Si/SiO$_{2}$ were acquired by FEI Helios NanoLab DualBeam 650. Atomic force microscopy height profiles of the samples deposited on Si/SiO$_{2}$ were measured in tapping mode by Bruker Innova in combination with v-type cantilever and SiN tips. Optical spectroscopy in reflection mode was performed with unpolarized light on SCs deposited on CaF$_{2}$ substrates, by Thermo Fisher FTIR spectrometer and Thermo Fisher Nicolet Continu$\mu$m microscope equipped with NO$_{2}$-cooled MCT and Si detectors, KBr and quartz beamsplitters, and a 15$\times$ (0.58 N.A.) Cassegrain objective. Surface enhanced Raman spectroscopy was carried out by Renishaw inVia micro-Raman microscope equipped with a 100$\times$ (0.85 N.A.) and 150$\times$ (0.95 N.A.) objectives, and 514 nm and 532 nm lasers at power 1.2 W/cm$^{2}$ and 0.1 W/cm$^{2}$, respectively, and integration time 1 s. All spectra were calibrated with respect to the first-order silicon LO phonon peak at 520 cm$^{-1}$ and recorded in backscattering geometry at room temperature. Raman measurements were performed on Au/G SCs and as a reference on bare graphene deposited on the same Si/SiO$_{2}$ substrate. Raman maps of graphene G and 2D, and BCB $\omega^{\star}=1655$ cm$^{-1}$ vibrational bands were scanned at 0.3 $\mu$m steps in both the directions in the plane of the sample. Renishaw WiRE 3.0 software was used to analyze the collected spectra, whose baseline was corrected to the third-order polynomial. Two- and four-probe electrical measurements in van der Pauw configuration were performed in vacuum ($10^{-6}$ mbar) by Keithley 2612A digital source-measure unit and Agilent 34410A digital voltmeter. In order to estimate the value of the residual carrier density $n_{0}\approx10^{11}$ cm$^{-2}$, two-probe resistance curves were fitted by the formula $R_{tot}=2R_{c}+R_{sh}$, with $R_{sh}=L/We\mu_{e}\left(n^{2}+n_{0}^{2}\right)^{1/2}$ the sheet resistance and $L$, $W$ the FET channel length and width, respectively. For four-probe resistance curves $R_{c}=0$ and $R_{sh}=4.53R_{tot}$, due to the van der Pauw geometrical correction. Carrier mobility was then evaluated by $\mu_{e}=\sigma/ne$. Photocurrent was measured using 0.80 eV, 1.46 eV, 1.94 eV, and 2.38 eV laser sources at different powers. The output power density was calibrated by a power meter. Responsivity was measured by using the following custom optical setup. An ASB-XE-175 source emits panchromatic light passing through a SP CM110 monochromator controlled by PC. Emerging monochromatic light is coupled to an optical fiber, modulated at 173 Hz by a mechanical chopper, and focused by a lens first on the samples, then on Si and Ge photodiodes. The AC current from the sample is measured through a current pre-amplifier (DL 1211) with a amplification factor $\mathcal{C}=10^{-6}$ A/V, whose output is read by a Signal Recovery 7265 lock-in amplifier. The photocurrent is given by the relation $J=2\pi\sqrt{2}V_{lock-in}*\mathcal{C}/4$, where the factor 2 is due to the peak-to-peak magnitude, the factor $\sqrt{2}$ originates from the lock-in amplifier rms amplitude, $\pi/4$ is the fundamental sine wave Fourier component of the square wave produced by the chopper, and $V_{lock-in}$ the amplitude of the signal read by the lock-in.

\begin{acknowledgments}
The authors acknowledge that this work has received funding from the European Union's Horizon 2020 research and innovation programme under grant agreement No. 696656 - GrapheneCore1.\\
\end{acknowledgments}
\section*{Author contribution}
\label{sec:AuthorContribution}
\noindent F.D.N., A.T., M.P., R.K., and V.P. conceived the experiments. V.M. and C.C. grew and transferred the graphene. N.P.P. fabricated the devices. F.D.N. performed the sample optical characterization, Raman spectroscopy, simulations, and data analysis. F.D.N., D.S., and N.P.P carried out the electrical measurements. All the authors discussed the experimental implementation, the results, and contributed in writing the paper.\\\\
The authors declare no competing financial interests.
\end{document}